\documentclass[smallextended]{svjour3}

\smartqed

\journalname{The European Physical Journal C}

\usepackage{amsmath}
\usepackage{xfrac}
\usepackage{booktabs}
\usepackage{graphicx}
\usepackage{dcolumn}
\usepackage{bm}
\usepackage{xcolor}
\usepackage{textcomp}

\newcommand{\sr}{\mathrm{sr}}
\newcommand{\ini}{\mathrm{ini}}
\newcommand{\en}{\mathrm{end}}

\newcommand{\sca}{\mathrm{S}}
\newcommand{\ten}{\mathrm{T}}

\newcommand{\D}{\mathrm{d}}

\begin{document}

\title{Numerical Study of Some Generalizations of the Starobinsky Inflationary Model}

\author{
Jordan Zambrano
\and \'Angel Rodr\'iguez
\and Melisa Alvear
\and Clara Rojas\thanks{\emph{Present address:} crojas@yachaytech.edu.ec}
}

\institute{
Jordan Zambrano \at
Universidad de Chile, Departamento de F\'isica, FCFM, Blanco Encalada 2008, Santiago, Chile
\and
\'Angel Rodr\'iguez, Melisa Alvear, Clara Rojas \at
Yachay Tech University, School of Physical Sciences and Nanotechnology, Hda. San Jos\'e S/N y Proyecto Yachay, 100119, Urcuqu\'i, Ecuador
}

\maketitle

\begin{abstract}
In this work, we perform a numerical study of three Starobinsky--type inflationary scenarios: the $\alpha$--Starobinsky inflationary model, the power--law Starobinsky inflationary model, and the power--law $\alpha$--Starobinsky inflationary model. For an appropriate choice of parameters, each scenario reproduces the standard Starobinsky limit. For each case, we derive the relevant slow--roll expressions in order to compute numerically the scalar and tensor power spectra over the corresponding parameter space and evaluate the associated inflationary observables. Finally, we provide a comparative analysis in the $(n_\sca,A_\sca)$ and $(r,n_\sca)$ parameter spaces using contour plots. Our results indicate that, for certain choices of parameters, the $\alpha$--Starobinsky model and the power--law $\alpha$--Starobinsky model are favored by \textit{Planck} 2018 observations.
\end{abstract}
\section{Introduction}

Inflation surged as a solution for some of the fine tuning puzzles within the hot big bang theory \cite{guth:1981}. Among them: the horizon problem, related to the homogeneous  observations of temperature anisotropies from the CMB $\left( \Delta T /T \sim 10^{-5}\right)$, the flatness problem, related to the observed value of curvature today, and the stability of this state throughout
the history of the Universe. Inflation is proposed as an era of rapid and exponential expansion of the universe at early times. This turns the fine tuning problems into causal consequences of inflationary dynamics \cite{linde:1982,guth:1985}.  Later, it was found that it also provides an explanation for the current large scale structure of the Universe. 

This theoretical framework is usually studied through a single scalar field called ``inflaton'' and a potential energy. The energy density of this field dominated the  expansion of the Universe during this short period of time, causing an accelerated expansion. In order to fulfill the requirements for accelerated expansion, inflaton energy density must have negative pressure, imposing some restrictions on the shape of the potential energy \cite{martin:2014}. The generation  of scalar quantum fluctuations from the scalar field \cite{mukhanov:1981} produces density perturbations that evolved into the temperature and polarization anisotropies in the CMB. Tensor perturbations to space--time produce primordial gravitational waves. 

In general, the inflationary potential is not known, and it must be sufficiently flat to meet the pressure condition. 
The most promising fact about inflation is that some of its predictions have already been observed and confirmed, such as the nearly scale--invariant scalar perturbations, its adiabatic nature, and the shape of the distribution of perturbations, which are in excellent agreement with CMB observation projects like WMAP \cite{jarosik:2011} and \textit{Planck} 2018 observations \cite{akrami:2020}.

Besides this success, there is still not a unique model that explains the mechanisms of inflation, and several have been proposed. Some of them are more favored by observations than others, and they are usually proposed from several areas of research, such as cosmology, quantum gravity, or string theory. 
The Starobinsky inflationary model is one of the inflationary scenarios most strongly supported by the \textit{Planck} 2018 data \cite{starobinsky:1980}. Recently, some generalizations of the Starobinsky inflationary model have been studied \cite{saini:2026a,saha:2025,saini:2023a,odintsov:2023,rodrigues:2023,meza:2021,canko:2020,chakravarty:2015,kallosh:2013b,kallosh:2013a,ellis:2013,roest:2014,cecotti:2014}, giving excellent agreement with the observational data.

In this work, we will focus on three Starobinsky--type classical inflation  models \cite{starobinsky:1980,noriji:2017}, one of the first proposed inflationary theories: $\alpha$–Starobinsky inflationary model \cite{saini:2025a}, power--law Starobinsky inflationary model \cite{saini:2023a},  and power--law $\alpha$--Starobinsky inflationary model \cite{saini:2026a}. Originally, Starobinsky inflation was proposed as a new kind of interaction of $(R + R^{2})$ in the Einstein--Hilbert action, with $R$ being the Ricci scalar. But it was also found that in the Einstein frame, it behaves as a flat potential for a scalar field \cite{kehagias:2014,ferrara}. This model offers good behavior compared with \textit{Planck} 2018 data and also provides an elegant passage to the radiation era through the reheating process \cite{akrami:2020,saini:2023a}.

$\alpha$--Starobinsky inflation model, power--law Starobinsky inflation model, and power--law $\alpha$--Starobinsky inflation model are generalizations of the Starobinsky inflation model. $\alpha$--Starobinsky inflation model considers the influence of an $\alpha$ parameter in the exponential. This model belongs to a greater class of superconformal models in supergravity known as $\alpha$--attractor \cite{saini:2025a}. This $\alpha$ parameter connects the various aspects of this model, in modified gravity, string theory, and PBH studies \cite{odintsov:2016,frolovsky:2025}. In particular, these models predict universal properties in observable diagrams like $n_\sca - r$; this includes the spectral index and tensor--to--scalar--ratio, and this behavior is dominated by the value of the selected $\alpha$ parameter \cite{kallosh:2013b,kallosh:2013a,ellis:2025,alho:2017}.

The power--law Starobinsky model was proposed as a generalization of $R^2$ gravity theory \cite{maeda:1989} in the form of $R^{\beta}$ interaction, and it returns to the classic Starobinsky model for $\beta = 2$. This kind of interaction has been widely studied in different areas, such as $f(R)$ theories in modified gravity and inflation as a generalization of the Starobinsky model \cite{saini:2023a,noriji:2011,muller:1990}. In this last area, it was shown in \cite{sebastiani:2014} that this model could be obtained from various potentials in the Einstein frame; it can also be included in a more general class of models called T--models. See \cite{cai:2014} for a deeper review.

When these two kinds of models are combined, the power--law $\alpha$-Starobinsky model is obtained, which depends on two parameters: $\alpha$ and $\beta$, when written in the Einstein frame \cite{saini:2025a}. This model includes both types of generalizations to the $R + R^2$ Starobinsky model. The generalization over $R^{\beta}$ interaction defines the power--law deviation from the standard model, and the generalization on $\alpha$, encoded in the conformal transformation, defines the relation with the K\"ahler potential \cite{saini:2026a}. Note that under fixed values for the parameters, we can recover the models described before.

This paper is organized in the following way: Section \ref{Background} presents the background equations and their form in the slow--roll approximation. In Section \ref{Perturbations}, the scalar and tensor perturbation equations are described. Later, in Section~\ref{Models}, three Starobinsky--type inflationary models are introduced, together with the Starobinsky inflationary model, to allow for a direct comparison.  In Section \ref{Contours}, our results are presented, where the contour plots are shown for each inflationary model. Finally, our  conclusions are presented in Section \ref{Conclusions}.

\section{Background Equations}
\label{Background}

The  equations of motion for the scalar field $\chi$ are given by \cite{liddle2000cosmological},

\begin{eqnarray}
\label{H2_ex}
H^2 &=& \dfrac{1}{3} \left[V(\chi)+\frac{1}{2}\dot\chi^2\right],\\
\label{chi_ex}
\ddot \chi &+& 3 H \dot{\chi} = - V_{,\chi},
\end{eqnarray}
where dots represent the derivation with respect to the physical time $t$, $H$ is the Hubble parameter, equal to  $H=\sfrac{\dot{a}}{a}$, with $a$ being the scale factor. The term  $V(\chi)$ is the inflationary potential of the scalar field, and $V_{,\chi}$ is the derivative of the inflationary potential with respect to the scalar field $\chi$.

Into the slow--roll approximation, the inflationary potential $V(\chi)$ dominates over the kinetic energy, so the background equations Eqs. \eqref{H2_ex} and   \eqref{chi_ex} reduce to,

\begin{eqnarray}
\label{H2_sr}
H^2 &\simeq& \dfrac{1}{3} V(\chi),\\
\label{a_sr}
3 H \dot\chi &\simeq& - V_{,\chi}.
\end{eqnarray}

\bigskip
The background equations Eqs. \eqref{H2_ex} and \eqref{chi_ex} are two coupled differential equations that must be integrated numerically using the Backward Differentiation Formula me\-thod of fifth order (BDF in Wolfram Mathematica$^\textsuperscript{\textregistered}$) \cite{Mathematica:2026},  with the step size selected adaptively by the solver during the integration process. The tolerance of the method is determined by \texttt{AccuracyGoal} and \texttt{PrecisionGoal}, which is set as ``Automatic''. In order to perform the numerical integration, we need the initial conditions: $a(0)=1$,  $\chi(0)=\chi_\ini$, and $\dot\chi(0)=\dot\chi_\sr(0)$, where we use the time derivative of the scalar field $\chi_\sr(t)$ as the initial condition, which is the solution to the equations of motion into the slow--roll approximation.

For early times, we calculate the slow--roll parameters that are given by \cite{liddle2000cosmological},

\begin{eqnarray}
\label{epsilon}
\epsilon &\simeq& \dfrac{1}{2} \left(\frac{V'}{V}\right)^2,\\
\label{eta}
\delta &\simeq& \dfrac{V''}{V},
\end{eqnarray}
where prime means the derivative with respect to the scalar field $\chi$.

In order to calculate the value of the scalar field at the end of inflation $\chi_\en$ we set $\epsilon=1$. To calculate the value of the scalar field at the beginning of inflation $\chi_\ini$, we need to calculate the number of e--folding using \cite{liddle2000cosmological},

\begin{equation}
\label{Ne}
N_e \simeq \int_{\chi_\en}^\chi \dfrac{V}{V'} \D \chi.
\end{equation}

Fixing the number of e--foldings $N_e$, we can calculate the initial value for the scalar field $\chi_\ini$.

\section{Perturbation Equations}
\label{Perturbations}

The scalar perturbations are described by the function $u=a\chi/\chi'$, where $\chi$ is a gauge--invariant variable corresponding to the Newtonian potential. The equation of motion of the scalar perturbation $u_k$ in Fourier space is given by \cite{mukhanov:1992},

\begin{equation}
u_k''+\left(k^2-\dfrac{z_{S}''}{z_{S}}\right)u_k=0,
\label{dotdotuk}
\end{equation}
where $z_{S}=a\chi'/\mathcal{H}$, $\mathcal{H}=a'/a$, and the prime indicates the derivative with respect to the conformal time $\eta$. The relation between $t$ and $\eta$ is given via the equation $\D t=a\,\D \eta$.

For tensor perturbations, we introduce the function $v_k=ah$, where $h$ represents the amplitude of the gravitational wave. Tensor perturbations obey a second--order differential equation analogous to Eq. (\ref{dotdotuk}), that is,

\begin{equation}
v_k''+\left(k^2-\dfrac{a''}{a}\right)v_k=0.
\label{dotdotvk}
\end{equation}
Considering the limits  $k^2\gg|z_{S}''/z_{S}|$ (short wavelength) and $k^2\ll|z_{S}''/z_{S}|$  (long wavelength), we find that  the  solutions to
Eqs. (\ref{dotdotuk})  exhibit the following asymptotic behavior:

\begin{equation}
\label{boundary_0}
u_k\rightarrow \frac{e^{-ik\eta}}{\sqrt{2k}}
\quad \left(k^2\gg|z_{S}''/z_{S}|, -k\eta\rightarrow \infty \right),
\end{equation}

\begin{equation}
\label{boundary_i} u_k\rightarrow A_k z  \quad \left(k^2\ll|z_{S}''/z_{S}|,-k\eta\rightarrow 0\right).
\end{equation}

\noindent Equation \eqref{boundary_0} is used as the initial condition for the perturbations. The same asymptotic conditions hold for tensor
perturbations. 
The selection of initial conditions is done following the standard Bunch--Davies vacuum condition, usually taken for quantum fluctuations during inflation. These are also part of an intense debate in the context of the trans--Planckian problem, where the duration of inflation has implications over the sizes that the scales we can observe now had in the early universe. See \cite{odintsov:2025a,odintsov:2025b,martin:2013} for more details.

The power spectra for scalar and tensor perturbations are given by the expressions

\begin{eqnarray}
\label{PS}
P_\sca(k)&=& \lim_{kt\rightarrow \infty} \frac{k^3}{2 \pi^2}\left|\dfrac{u_k(t)}{z_{S}(t)} \right|^2,\\
\label{PT}
P_\ten(k)&=& \lim_{kt\rightarrow \infty} \frac{k^3}{2 \pi^2}\left|\dfrac{v_k(t)}{a(t)} \right|^2,
\end{eqnarray}
from Eq. \eqref{PS}, we can calculate the scalar spectral index, defined by,

\begin{equation}
\label{nS}
n_\sca(k)= 1+\frac{\D\ln P_\sca(k)}{\D\ln k}.
\end{equation}

\bigskip
\noindent 
In addition, the tensor--to--scalar ratio $r(k)$ is defined as \cite{habib:2005b},

\begin{equation}
\label{R}
r(k)=8\,\frac{P_\ten(k)}{P_\sca(k)}.
\end{equation}

In Section~\ref{Background}, we derived the scale factor $a(t)$ and the scalar field $\chi(t)$ in terms of the physical time $t$, instead of the conformal time $\eta$. Following the same approach, it is necessary to express the equations for the scalar and tensor perturbations, Eqs. \eqref{dotdotuk} and \eqref{dotdotvk}, in terms of the physical time $t$ as well, so that they can be integrated numerically. This is explained in detail by Zambrano \textit{et al.} \cite{zambrano:2026}. Working in physical time leads to the requirement that the initial conditions, Eqs. \eqref{boundary_0} and \eqref{boundary_i} be evaluated at $+ k t \rightarrow 0$ and for $+ k t \rightarrow \infty$, respectively. This change in the evaluation of the initial conditions leads to a change in the behavior of the mode in each limit, now the mode is going to be oscillatory in $+ kt \rightarrow \infty$ and freezes for $+ k t \rightarrow 0$.
Finally, in order to find the scalar spectral index $n_\sca(k)$ and $r(k)$, we implement the fit of the scalar  $P_\sca(k)$  and tensor power spectra $P_\ten(k)$ as proposed in  \textit{Planck} articles \cite{giare:2023b,das:2023,vazquez:2020,finelli:2018,vazquez:2013}.

\section{Models}
\label{Models}

\subsection{Starobinsky Inflationary Model}

The Starobinsky inflationary model was proposed for A. A. Starobinsky in  $1980$ \cite{starobinsky:1980}. This model is among the most favored by the \textit{Planck} 2018 results and is described by the following potential $V(\chi)$, 

\begin{equation}
\label{VS}
V(\chi)=\dfrac{3}{4} M^2 \left(1-e^{-\sqrt{\frac{2}{3}}\chi}\right)^2,
\end{equation}
where $M$ is a parameter to be normalized by the predicted amplitude of the scalar power spectrum $A_\sca$ at the pivot scale $k_*=0.05$ Mpc$^{-1}$  to the \textit{Planck} 2018 results.

The slow--roll parameters are given by,

\begin{eqnarray}
\label{epsilonS}
\epsilon &=& \dfrac{4}{3} \dfrac{1}{\left(e^{\sqrt{\frac{2}{3}}\chi}-1\right)^2},\\
\label{deltaS}
\delta &=& -\dfrac{4}{3} \dfrac{\left(e^{\sqrt{\frac{2}{3}}\chi}-2\right)}{\left(e^{\sqrt{\frac{2}{3}}\chi}-1\right)^2}.
\end{eqnarray}

Considering the end of inflation when $\epsilon=1$ the end value for the scalar field $\chi$ is obtained,

\begin{equation}
\label{chiendS}
\chi_\en=\sqrt{\dfrac{3}{2}} \ln \left(1+\dfrac{2}{\sqrt{3}} \right),
\end{equation}
then using the number of e--folds $N_e$ we can compute the initial value for the scalar field $\chi_\ini$,

\begin{eqnarray}
\label{chiiniS}
\nonumber
\chi_\ini=&-&2\,\sqrt{\dfrac{2}{3}} N_e-\sqrt{\dfrac{3}{2}}e^{\sqrt{\frac{2}{3}}\chi_\en}+\chi_\en,\\
\nonumber
&-& \sqrt{\dfrac{3}{2}} W_{-1} \left(-e^{-\frac{4}{3} N_e+\sqrt{\frac{2}{3}}\chi_\en-e^{\sqrt{\frac{2}{3}}\chi_\en}}\right),\\
\end{eqnarray}
where $W_{-1}$ is the $-1$ branch of the Lambert $W$ function.
 
For our numerical calculation, we need the solution for the slow--roll expression for the scale factor $a(t)$ and the scalar field $\chi(t)$, which are given by

\begin{eqnarray}
\nonumber
\ln a(t)&=&\dfrac{1}{2} M t - \dfrac{3}{4} \ln\left(e^{\sqrt{\frac{2}{3}}\chi_\ini}\right)+\dfrac{3}{4}\ln\left(e^{\sqrt{\frac{2}{3}}\chi_\ini} - \dfrac{2M}{3}t\right),\\
\nonumber\\\\
\chi(t) &=& \sqrt{\dfrac{3}{2}}\ln\left( -\dfrac{2}{3} Mt+e^{\sqrt{\frac{2}{3}}\chi_\ini}\right).
\end{eqnarray}

\subsection{$\alpha$--Starobinsky Inflationary Model}

The $\alpha$--Starobinsky inflationary model is also known as the $E$--model and was originally introduced in the context of $\alpha$--attractors by Kallosh, Linde, and Roest, as well as by Cecotti \cite{kallosh:2013a,kallosh:2013b,roest:2014,cecotti:2014}. It can be viewed as an $f(R)$ deformation of the Starobinsky scenario, and it belongs to the $\alpha$--attractors inflationary scenarios. In recent years, this framework has attracted considerable attention, particularly due to the tight constraints imposed by the \textit{Planck} 2018 observations \cite{saini:2025a,haque:2025,frolovsky:2025,ellis:2025,saha:2025}.

The corresponding potential $V(\chi)$ is described by the following expression, 

\begin{equation}
\label{VaS}
V(\chi)=\dfrac{3}{4} M^2 \left(1-e^{-\sqrt{\frac{2}{3\alpha}}\chi}\right)^2,
\end{equation}
where $\alpha$ is known as the K\"ahler curvature parameter, and for $\alpha=1$  the Starobinsky inflationary model, it is recovered. 

The slow--roll parameters are given by \cite{saha:2025},

\begin{eqnarray}
\label{epsilonaS}
\epsilon &=& \dfrac{4}{3\,\alpha} \frac{1}{\left(e^{\sqrt{\frac{2}{3\alpha}}\chi}-1\right)^2},\\
\label{deltaaS}
\delta &=& -\dfrac{4}{3\,\alpha} \dfrac{\left(e^{\sqrt{\frac{2}{3\alpha}}\chi}-2\right)}{\left(e^{\sqrt{\frac{2}{3\alpha}}\chi}-1\right)^2}.
\end{eqnarray}

Considering the end of inflation, that means $\epsilon=1$ we obtain the  end value for the scalar field $\chi_\en$,

\begin{equation}
\label{chiendaS}
\chi_\en=\sqrt{\dfrac{3\alpha}{2}} \ln \left(1+\dfrac{2}{\sqrt{3\alpha}} \right),
\end{equation}
then using the number of e--folds $N_e$ we can compute the initial value for the scalar field $\chi_\ini$,

\begin{eqnarray}
\label{chiiniaS}
\nonumber
\chi_\ini=&-&2\,\sqrt{\dfrac{2}{3\alpha}} N_e-\sqrt{\dfrac{3 \alpha}{2}}e^{\sqrt{\frac{2}{3 \alpha}}\chi_\en}+\chi_\en\\
\nonumber
&-& \sqrt{\dfrac{3 \alpha}{2}}\, W_{-1} \left(-e^{-\frac{4}{3 \alpha} N_e+\sqrt{\frac{2}{3\alpha}}\chi_\en-e^{\sqrt{\frac{2}{3\alpha}}\chi_\en}}\right),\\
\end{eqnarray}
where $W_{-1}$ is the $-1$ branch of the Lambert $W$ function.

In order to perform the numerical calculation, the solutions of the slow--roll equations for the scale factor $a(t)$ and the scalar field $\chi(t)$ are needed, which are given by,

\begin{eqnarray}
\nonumber
\ln a(t)&=&\dfrac{1}{2} M t  - \dfrac{3\,\alpha}{4} \ln\left(e^{\sqrt{\frac{2}{3\,\alpha}}\chi_\ini}\right)\\
        & +&  \dfrac{3\,\alpha}{4}\ln\left(e^{\sqrt{\frac{2}{3\,\alpha}}\chi_\ini} - \frac{2M}{3\,\alpha}t\right),\\
\chi(t) &=& \sqrt{\dfrac{3\,\alpha}{2}}\ln\left( -\dfrac{2}{3\,\alpha} Mt+e^{\sqrt{\frac{2}{3\,\alpha}}\chi_\ini}\right).
\end{eqnarray}

\subsection{Power--law Starobinsky Inflationary Model}

The power--law Starobinsky inflationary model can be formulated within the class of $T$--models \cite{saini:2023a,odintsov:2023,chakravarty:2015,kallosh:2013b}. It corresponds to a power--law  $f(R)$ deformation of the original Starobinsky scenario \cite{burikham:2024,saini:2023a,codello:2015,ben:2014}.  This scenario is  tightly constrained by the \textit{Planck}  2018 observations only for a few of the chosen parameters.

The associated Einstein--frame potential $V(\chi)$ is given by,

\begin{eqnarray}
\label{VplaS}
\nonumber
V(\chi)&=&\left(\dfrac{\beta-1}{2}\right)\left(\frac{6M^2}{\beta^\beta}\right)^{\frac{1}{\beta-1}}e^{\sqrt{\frac{2}{3}}\left(\frac{2-\beta}{\beta-1}\right)\chi}\\
&\times& \left(1-e^{-\sqrt{\frac{2}{3}}\chi}\right)^{\frac{\beta}{\beta-1}}
\end{eqnarray}
where $\beta$ represents the deformation from the Starobinsky inflationary model. For $\beta=2$ the Starobinsky inflationary model is recovered.

The slow--roll parameters are given by,

\begin{eqnarray}
\label{epsilonplS}
\epsilon &=& \dfrac{1}{3}\dfrac{\left[2-2\,\beta +\left( \beta-2\right)e^{\sqrt{\frac{2}{3}}\chi}\right]^2}{\left(\beta-1\right)^2\left( e^{\sqrt{\frac{2}{3}}\chi}-1\right)^2},\\
\label{deltaplS}
\nonumber
\delta &=& \dfrac{2}{3} \dfrac{\left[4 \left(\beta-1\right)^2 -\left(\beta-1\right) \left(5\,\beta-8\right) e^{\sqrt{\frac{2}{3}}\chi}+\left(\beta-2\right)^2 e^{2\sqrt{\frac{2}{3}}\chi}\right]}{\left(\beta-1\right)^2\left( e^{\sqrt{\frac{2}{3}}\chi}-1\right)^2}.\\
\end{eqnarray}

Considering the end of inflation, that means $\epsilon=1$ we obtain the  end value for the scalar field $\chi_\en$,

\begin{equation}
\label{chiendplS}
\chi_\en=\sqrt{\dfrac{3}{2}} \ln\left[\dfrac{\left(\beta-1\right)\left( 1+\beta+\sqrt{3}\,\beta\right)}{2\,\beta\left(\beta-1\right)-1} \right],
\end{equation}
then, into the slow--roll approximation, the number of e--folds $N_e$ can be used to determine the initial value of the scalar field, $\chi_{\ini}$ for each value of $\beta$. This model does not have an analytical solution for $a_\sr(t)$ and $\chi_\sr(t)$; therefore, it is necessary to solve Eqs. \eqref{H2_sr} and \eqref{a_sr} numerically.

\subsection{Power--law $\alpha$--Starobinsky Inflationary Model}

The power--law $\alpha$--Starobinsky model arises from combining the $\alpha$--Starobinsky (or $\alpha$--attractor) framework with a power--law $f(R)$ deformation \cite{saini:2026a,ellis:2013}. Consequently, it is characterized by the parameter $\alpha$ of $\alpha$--Starobinsky inflation, together with the exponent $\beta$ that controls the power--law modification.  This scenario is strongly favored by the constraints from the \textit{Planck}  2018  observations.

The corresponding Einstein--frame potential $V(\chi)$ is given by,

\begin{eqnarray}
\label{VplS}
\nonumber
V(\chi)&=&\left(\dfrac{\beta-1}{2}\right)\left(\dfrac{6M^2}{\beta^\beta}\right)^{\frac{1}{\beta-1}}e^{\sqrt{\frac{2}{3\alpha}}\left(\frac{2-\beta}{\beta-1}\right)\chi}\\
&\times& \left(1-e^{-\sqrt{\frac{2}{3\alpha}}\chi}\right)^{\frac{\beta}{\beta-1}},
\end{eqnarray}
this potential reduces to the Starobinsky inflationary model in the Einstein frame for $\beta=2$ and $\alpha=1$.

The slow--roll parameters are given by,

\begin{align}
\label{epsilonplaS}
\epsilon &=\dfrac{1}{3\alpha}\frac{\left[2-2\,\beta +\left( \beta-2\right)e^{\sqrt{\frac{2}{3\alpha}}\chi}\right]^2}{\left(\beta-1\right)^2\left( e^{\sqrt{\frac{2}{3\alpha}}\chi}-1\right)^2},\\
\nonumber
\label{deltaplaS}
\delta &= \dfrac{2}{3\alpha} \left\{\frac{\left[4 \left(\beta-1\right)^2 -\left(\beta-1\right) \left(5\,\beta-8\right) e^{\sqrt{\frac{2}{3\alpha}}\chi}\right.}{\left(\beta-1\right)^2\left( e^{\sqrt{\frac{2}{3\alpha}}\chi}-1\right)^2}\right.\\
&+  \left. \frac{\left.\left(\beta-2\right)^2 e^{2\sqrt{\frac{2}{3\alpha}}\chi}\right]}{\left(\beta-1\right)^2\left( e^{\sqrt{\frac{2}{3\alpha}}\chi}-1\right)^2}\right\}.
\end{align}

Considering the end of inflation, that means $\epsilon=1$ we obtain the  end value for the scalar field $\chi_\en$,

\begin{eqnarray}
\label{chiendplaS}
\nonumber
\chi_\en =\sqrt{\dfrac{3\,\alpha}{2}} &\ln\left[\dfrac{-4+3\,\alpha\left(\beta-1\right)^2-2\left(\beta-3\right)\beta}{3\,\alpha\left(\beta-1\right)^2-\left(\beta-2\right)^2} \right.\\
&+ \left. \dfrac{\sqrt{3\,\alpha}\,\beta\left(\beta-1\right)}{3\,\alpha\left(\beta-1\right)^2-\left(\beta-2\right)^2} \right].
\end{eqnarray}
Then, using the number of e--folds $N_e$, we can compute numerically the initial value for the scalar field $\chi_\ini$; which does not have  an analytical solution for this model.
In this inflationary model, we also do not have an analytical solution for $a_\sr(t)$ and $\chi_\sr(t)$, so we have to solve Eqs. \eqref{H2_sr}  and \eqref{a_sr} numerically as well.

\section{Parameters Space}
\label{Contours}

In this section, we present two--dimensional contour plots  of each inflationary model through the $(n_\sca,A_\sca)$ and $( n_\sca,r)$ plane, showing the $95\,\%$ and $68\,\%$ credible regions from \textit{Planck} 2018 data, and overlaying the model predictions for $N_e=50$ and $N_e=60$. Each point was obtained by numerically integrating the background equations together with the perturbation equations.

\subsection{Starobinsky Inflationary Model}

\begin{figure}[th!]
\centering
\includegraphics[scale=0.5]{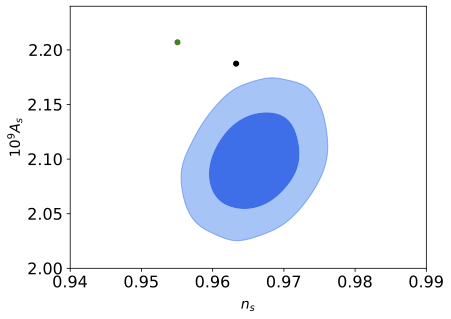}
\caption{The contour plot $(n_\sca,A_\sca)$ for the Starobinsky inflationary model for $N_e=50$ (green point) and $N_e=60$  (black point). In this Figure it can be observed that the points for $N_e=50$ and $N_e=60$ are far from the Planck contour.}
\label{contour_ASnSS}
\end{figure}

\begin{figure}[th!]
\centering
\includegraphics[scale=0.5]{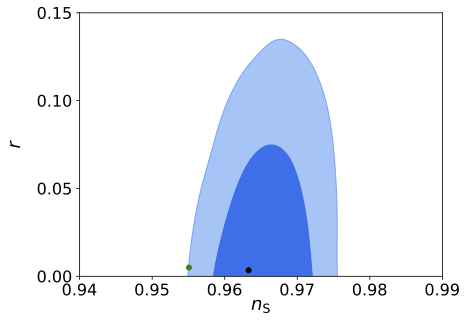}
\caption{The contour plot $(n_\sca,r)$ for the Starobinsky inflationary model for  $N_e=50$ (green point) and $N_e=60$  (black point).}
\label{contour_rnS}
\end{figure}

In this section, we present the results for the Starobinsky inflationary model \cite{starobinsky:1980} in order to compare them with those of the generalized models. Figs. \ref{contour_ASnSS} and \ref{contour_rnS} show the contour plots for $N_e=50$ and $N_e=60$, where it can be observed that the predicted values of the observables lie within the $95\,\%$ and $68\,\%$ confidence levels. These values are summarized in Tables \ref{observables_ASnSSN50} and \ref{observables_ASnSSN60}.

\begin{table}[th]
\centering
\begin{tabular}{ccc}
\toprule
\cmidrule(lr){1-3}
$n_{\sca_{0.05}}$ & $10^9 A_{\sca_{0.05}}$ & $r_{\sca_{0.002}}$  \\
\midrule
0.955062 & 2.20698 & 0.00499007 \\
\bottomrule
\end{tabular}
\caption{Observables $n_{\sca_{0.05}}$, $10^9 A_{\sca_{0.05}}$ and $r_{\sca_{0.002}}$ with respect to the values reported by \textit{Planck} 2018 results for the Starobinsky inflationary model, calculated numerically considering $N_e=50$.}
\label{observables_ASnSSN50}
\end{table}

\begin{table}[th]
\centering
\begin{tabular}{ccc}
\toprule
$n_{\sca_{0.05}}$ & $10^9 A_{\sca_{0.05}}$ & $r_{\sca_{0.002}}$  \\
\midrule
0.963295 & 2.18744& 0.00345018  \\
\bottomrule
\end{tabular}
\caption{Observables $n_{\sca_{0.05}}$, $10^9 A_{\sca_{0.05}}$ and $r_{\sca_{0.002}}$ with respect to the values reported by \textit{Planck} 2018 results for the Starobinsky inflationary model, calculated numerically considering $N_e=60$.}
\label{observables_ASnSSN60}
\end{table}

\begin{table}[htbp!]
\centering
\begin{tabular}{cccc}
\toprule
\multicolumn{2}{c}{$N_e=50$} & \multicolumn{2}{c}{$N_e=60$} \\
\cmidrule(lr){1-2}\cmidrule(lr){3-4}
$t_{\ini_{0.05}}$ & $\eta_{\ini_{0.05}}$ 
& $t_{\ini_{0.05}}$ & $\eta_{\ini_{0.05}}$ \\
\cmidrule(lr){1-2}\cmidrule(lr){3-4}
240347 & $-126360.0$ & 325197 & $-154587.9$ \\
\bottomrule
\end{tabular}
\caption{Initial conditions to the numerical calculation of $A_{\sca_{0.05}}$ considering $N_e=50$ and $N_e=60$ for the Starobinsky inflationary model.}
\label{S_initial_ASNe60}
\end{table}

The initial conditions  in terms of cosmic time and conformal time  used for the numerical calculation of $A_{\sca_{0.05}}$ for $N_e=50$ and $N_e=60$, in the Starobinsky inflationary model, are shown in Table \ref{S_initial_ASNe60}.

\subsection{$\alpha$--Starobinsky Inflationary Model}

In this subsection, we present our results for the $\alpha$ --Starobinsky inflationary model. We adopt the parameter values of $\alpha$ considered by Saini and Nautiyal \cite{saini:2025a}, namely $\alpha = \{0.01,\ 1,\ 20,\ 50,\ 100\}$.

\begin{figure}[th!]
\centering
\includegraphics[scale=0.5]{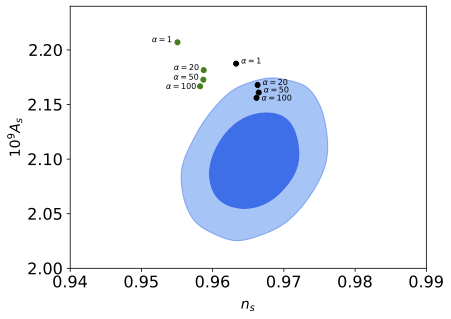}
\caption{The contour plot $(n_\sca,A_\sca)$ for the $\alpha$--Starobinsky inflationary model for $N_e=50$ (green points) and $N_e=60$  (black points). In this Figure it can be observed that all points for $N_e=50$ are far from the Planck contour, and for $N_e=60$ the point $\alpha=1$ has the same behavior.}
\label{contour_ASnSalphaS}
\end{figure}

\begin{figure}[th!]
\centering
\includegraphics[scale=0.5]{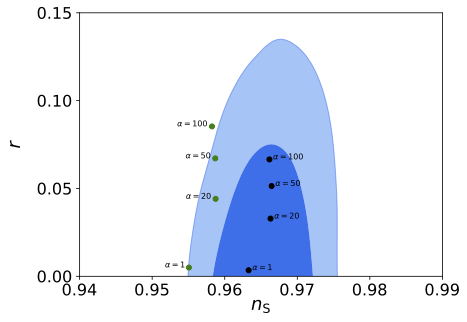}
\caption{The contour plot $(n_\sca,r)$ for the $\alpha$--Starobinsky inflationary model for  $N_e=50$ (green points) and $N_e=60$  (black points). In this Figure it can be observed that the point $\alpha=100$ for $N_e=50$ is outside from the Planck contour.}
\label{contour_rnSalphaS}
\end{figure}

Figure \ref{contour_ASnSalphaS} shows the $(n_\sca,A_\sca)$ parameter space for $N_e=50$ and $N_e=60$, consistent with the predictions of Saini and Nautiyal \cite{saini:2025a}. We find that four of the points computed for $N_e=60$ lie within the $68\,\%$ confidence level allowed by the \textit{Planck} 2018 constraints. Figure \ref{contour_rnSalphaS} displays the $(n_\sca, r)$ plane for the same values of $N_e$. In this case, four of the $N_e=60$ points fall within the \textit{Planck} $68\,\%$ confidence level, indicating that the model is favored by current CMB measurements. The numerical values corresponding to these points are summarized in Tables \ref{observables_ASnSaS} and \ref{observables_nSraS}.

\begin{table}[htbp]
\begin{center}
\begin{tabular}{ccccc}
\toprule
 & \multicolumn{2}{c}{$N_e=50$} & \multicolumn{2}{c}{$N_e=60$} \\
\cmidrule(lr){2-3}\cmidrule(lr){4-5}
$\alpha$ 
& $n_{\sca_{0.05}}$ & $10^9 A_{\sca_{0.05}}$  
& $n_{\sca_{0.05}}$ & $10^9 A_{\sca_{0.05}}$  \\
\midrule
0.01 & 0.886614 & 2.39069 & 0.886593 & 2.39058  \\
1    & 0.955062 & 2.20698 & 0.963295 & 2.18744  \\
20   & 0.958747 & 2.18151 & 0.966313 & 2.16788 \\
50   & 0.958696 & 2.17274 & 0.966464 & 2.16099 \\
100  & 0.958324 & 2.16673 & 0.966158 & 2.15612\\
\bottomrule
\end{tabular}
\caption{Observables  $n_{\sca_{0.05}}$ and $10^9 A_{\sca_{0.05}}$ respect to the value reported by \textit{Planck} 2018 results for the $\alpha$--Starobinsky inflationary model   calculated  numerically for five different values of the parameter $\alpha$ considering $N_e=50$, and $N_e=60$.}
\label{observables_ASnSaS}
\end{center}
\end{table}

\begin{table}[htbp]
\begin{center}
\begin{tabular}{ccccc}
\toprule
 & \multicolumn{2}{c}{$N_e=50$} & \multicolumn{2}{c}{$N_e=60$} \\
\cmidrule(lr){2-3}\cmidrule(lr){4-5}
$\alpha$ 
& $n_{\sca_{0.05}}$ & $r_{0.002}$ 
& $n_{\sca_{0.05}}$ & $r_{0.002}$ \\
\midrule
0.01 & 0.886614 & 0.000280065 & 0.886593 & 0.00028031  \\
1    & 0.955062 & 0.00499007 & 0.963295 & 0.00345018  \\
20   & 0.958747 & 0.0441007 & 0.966313 & 0.0328602 \\
50   & 0.958696 & 0.0671406  & 0.966464 & 0.051403 \\
100  & 0.958324 & 0.0853467 & 0.966158 & 0.0665394\\
\bottomrule
\end{tabular}
\caption{Observables $n_{\sca_{0.05}}$ and $r_{0.002}$ respect to the value reported by \textit{Planck} 2018 results for the $\alpha$--Starobinsky inflationary model   calculated  numerically for five different values of the parameter $\alpha$ considering $N_e=50$, and $N_e=60$.}
\label{observables_nSraS}
\end{center}
\end{table}

These results are consistent with those reported by Saini and Nautiyal \cite{saini:2025a}, where they found that using MCMC $\alpha = 1.0^{+39.81}_{-2.5\times 10^{-6}}$ resulted in a low level of constraining power for the data they used. Another study showed constraints for $\alpha$ in the range $\alpha = 7.56^{+5.15}_{-5.15}$, using data from Planck 2018, BAO data from BOSS, SDSS, 6dF Galaxy Survey, among others. See \cite{rodrigues2021observational} for more details on the MCMC analysis they conducted.

The initial conditions  in terms of cosmic time and conformal time  used for the numerical calculation of $A_{\sca_{0.05}}$ for $N_e=50$ and $N_e=60$ in the $\alpha$--Starobinsky inflationary model, are shown in Table \ref{alphaS_initial_ASNe50} and Table \ref{alphaS_initial_ASNe60},

\begin{table}[htbp]
\centering
\begin{tabular}{ccccc}
\toprule
 & \multicolumn{2}{c}{$N_e=50$} \\
\cmidrule(lr){2-3}
$\alpha$ 
& $t_{\ini_{0.05}}$ & $\eta_{\ini_{0.05}}$    \\
\midrule
0.01 & $1.4957 \times 10^6$ & $ -466326.6$   \\
1    & $240347$ & $-126360.0$   \\
20   & $34778.9$ & $-42629.4$&  \\
50   & $20785.9$ & $ -34464.9$&   \\
100  & $14644.5$ & $-30487.5$ &  \\
\bottomrule
\end{tabular}
\caption{Initial conditions to the numerical calculation of $A_{\sca_{0.05}}$ considering $N_e=50$ for the $\alpha$--Starobinsky inflationary model.}
\label{alphaS_initial_ASNe50}
\end{table}

\begin{table}[htbp]
\centering
\begin{tabular}{ccccc}
\toprule
 & \multicolumn{2}{c}{$N_e=60$} \\
\cmidrule(lr){2-3}
$\alpha$ 
& $t_{\ini_{0.05}}$ & $\eta_{\ini_{0.05}}$    \\
\midrule
0.01 & $1.4957 \times 10^6$ & $ -466326.6$   \\
1    & $325197$ & $ -154587.9$   \\
20   & $49197.3$ & $-50216.2$&  \\
50   & $30210.8$ & $ -40070.3$&   \\
100  & $21880.2$ & $-35142.3$ &  \\
\bottomrule
\end{tabular}
\caption{Initial conditions to the numerical calculation of $A_{\sca_{0.05}}$ considering $N_e=60$ for the $\alpha$--Starobinsky inflationary model.}
\label{alphaS_initial_ASNe60}
\end{table}

\subsection{Power--law Starobinsky Inflationary Model}

In this subsection, we present our results for the power--law Starobinsky inflationary model. We adopt the parameter values of $\beta$ considered by Saini and Nautiyal \cite{saini:2023a}, namely $\beta = \{1.90,\ 1.95,\ 2.00,\ 2.05,\ 2.10\}$.

\begin{figure}[th!]
\centering
\includegraphics[scale=0.5]{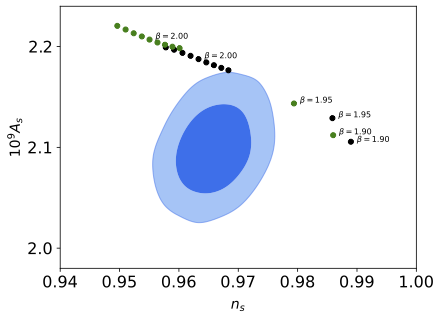}
\caption{The contour plot $(n_\sca,A_\sca,)$  for the power-- law Starobinsky inflationary model for and $N_e=50$ (green points) and $N_e=60$  (black points). In this Figure it can be observed that the  points $\beta=\left\{1.90, 1.95, 2.00 \right\}$ for $N_e=50$ and for $N_e=60$ live far from the Planck contour.}
\label{contours_ASnSplS}
\end{figure}

\begin{figure}[th!]
\centering
\includegraphics[scale=0.5]{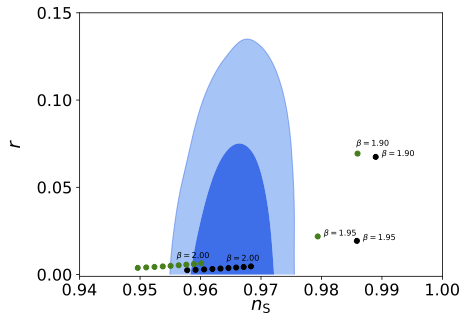}
\caption{The contour plot $(n_\sca,r)$ for the power law Starobinsky inflationary model for  $N_e=50$ (green points) and $N_e=60$  (black points). In this Figure it can be observed that the  points $\beta=\left\{1.90, 1.95 \right\}$ for $N_e=50$ and for $N_e=60$ live far from the Planck contour.}
\label{contours_rnSplS}
\end{figure}

Figure \ref{contours_ASnSplS} displays the $(n_\sca,A_\sca,)$ parameter space for $N_e=50$ and $N_e=60$, consistent with the predictions by Saini and Nautiyal \cite{saini:2023a}. We observe that none of the points computed for either $N_e=50$ or $N_e=60$ fall within the $95\,\%$ or $68\,\%$ confidence levels allowed by the \textit{Planck} 2018 constraints. In contrast, Fig. \ref{contours_rnSplS} shows the $(n_\sca,r)$ plane for the same values of $N_e$. In this case, only the $\beta=2$ point lies within the \textit{Planck} $68\,\%$ confidence level, for both $N_e=50$ and $N_e=60$. Therefore, within the set of values explored so far, this model is disfavored by observations for values different from $\beta=2$ as proposed by Saini and Nautiyal \cite{saini:2023a}. The numerical values corresponding to these points are summarized in Tables \ref{observables_ASnSplS} and \ref{observables_nSrplS}. In Table \ref{observables_nSrplS}, we find that our results are in good agreement with the combined ACT and DESI data, which favor slightly higher values of $n_{\sca_{0.05}}$ and $r_{0.002}$ \cite{adame:2025,oikonomou:2025,odintsov:2025c}. In references \cite{oikonomou:2025,odintsov:2025c}, an approach is offered in which the power--law Starobinsky model is treated in the Jordan frame; these results are contrasted with Planck and ACT data and obtain consistent results with the ones obtained in this work.

\begin{table}[th!]
\begin{center}
\begin{tabular}{ccccc}
\toprule
 & \multicolumn{2}{c}{$N_e=50$} & \multicolumn{2}{c}{$N_e=60$} \\
\cmidrule(lr){2-3}\cmidrule(lr){4-5}
$\beta$ 
& $n_{\sca_{0.05}}$ & $10^9 A_{\sca_{0.05}}$  
& $n_{\sca_{0.05}}$ & $10^9 A_{\sca_{0.05}}$  \\
\midrule
1.90 & 0.985964 & 2.11210 & 0.988974 & 2.10556  \\
1.95 & 0.979380 & 2.14356 & 0.985859 & 2.12899  \\
2.00 & 0.955049 & 2.20693 & 0.963302 & 2.18756 \\
2.05 & 0.915345 & 2.30624  & 0.922059 & 2.28909 \\
2.1  & 0.866949 & 2.43416 & 0.871087 & 2.44480\\
\bottomrule
\end{tabular}
\caption{Observables $n_{\sca_{0.05}}$ and $10^9 A_{\sca_{0.05}}$   respect to the value reported by \textit{Planck} 2018 results for the power law Starobinsky inflationary model  calculated  numerically for five different values of the parameter $\beta$ considering $N_e=50$, and $N_e=60$.}
\label{observables_ASnSplS}
\end{center}
\end{table}

\begin{table}[th!]
\begin{center}
\begin{tabular}{ccccc}
\toprule
 & \multicolumn{2}{c}{$N_e=50$} & \multicolumn{2}{c}{$N_e=60$} \\
\cmidrule(lr){2-3}\cmidrule(lr){4-5}
$\beta$ 
& $n_{\sca_{0.05}}$ & $r_{0.002}$ 
& $n_{\sca_{0.05}}$ & $r_{0.002}$ \\
\midrule
1.90 & 0.985964 & 0.0693332 & 0.988974 & 0.0674359  \\
1.95 & 0.979380 & 0.0218554 & 0.985859 & 0.0193012  \\
2.00 & 0.955049 & 0.00498975 & 0.963302 & 0.00345025 \\
2.05 & 0.915345 & 0.000871498  & 0.922059 & 0.000404467 \\
2.1  & 0.866949 & 0.000131801 & 0.871087 & 0.0000380349\\
\bottomrule
\end{tabular}
\caption{Observables $n_{\sca_{0.05}}$ and $r_{0.002}$ respect to the value reported by \textit{Planck} 2018 results for the power law Starobinsky inflationary model    calculated  numerically for five different values of the parameter $\beta$ considering $N_e=50$, and $N_e=60$.}
\label{observables_nSrplS}
\end{center}
\end{table}

Based on a recent studied conducted by Meza \cite{meza:2021} \textit{et. al}, which also explored a generalization of the Starobinsky inflationary model, using parameter $p$, if equating the exponents of both models,

\begin{equation}
\frac{2 p}{2 p -1}=\frac{\beta}{\beta-1} \rightarrow \beta=2p,
\end{equation}
then the new set of values to consider is 
$\beta=\{1.992,1.994,\\
1.996,1.998,2.002,2.004,2.006,2.008\}$. Under these values, the model is highly favored by the \textit{Planck} 2018 data and is also consistent with the values of $\beta$ obtained from the MCMC analysis of Saini and Nautiyal \cite{saini:2023a}, where $\beta=1.966^{+0.035}_{-0.042}$ is within $95\,\%$ C.L.   The numerical values corresponding to these points are summarized in Tables \ref{observables_ASnSplS_additional} and \ref{observables_nSrplS_additional}.

\begin{table}[th!]
\begin{center}
\begin{tabular}{ccccc}
\toprule
 & \multicolumn{2}{c}{$N_e=50$} & \multicolumn{2}{c}{$N_e=60$} \\
\cmidrule(lr){2-3}\cmidrule(lr){4-5}
$\beta$ 
& $n_{\sca_{0.05}}$ & $10^9 A_{\sca_{0.05}}$  
& $n_{\sca_{0.05}}$ & $10^9 A_{\sca_{0.05}}$  \\
\midrule
1.992 & 0.960119 & 2.19846 & 0.968337 & 2.17648  \\
1.994 & 0.958908 & 2.19986 & 0.967138 & 2.17886  \\
1.996 & 0.957654 & 2.20183 & 0.965896 & 2.18154 \\
1.998 & 0.956388 & 2.20405  & 0.964616 & 2.18430 \\
2.002  & 0.95375 & 2.21005 & 0.961962 & 2.19081\\
2.004  & 0.952383 & 2.21321 & 0.960594 & 2.19374\\
2.006  & 0.951017 & 2.21691 & 0.959188 & 2.19684\\
2.008  & 0.949616 & 2.22054 & 0.957789 & 2.19920\\
\bottomrule
\end{tabular}
\caption{Observables $n_{\sca_{0.05}}$  and $10^9 A_{\sca_{0.05}}$  respect to the value reported by \textit{Planck} 2018 results for the power law Starobinsky inflationary model calculated  numerically for  values of the parameter $\beta$ close to $\beta=2$, considering $N_e=50$, and $N_e=60$.}
\label{observables_ASnSplS_additional}
\end{center}
\end{table}

\begin{table}[th!]
\begin{center}
\begin{tabular}{ccccc}
\toprule
 & \multicolumn{2}{c}{$N_e=50$} & \multicolumn{2}{c}{$N_e=60$} \\
\cmidrule(lr){2-3}\cmidrule(lr){4-5}
$\beta$ 
& $n_{\sca_{0.05}}$ & $r_{0.002}$ 
& $n_{\sca_{0.05}}$ & $r_{0.002}$ \\
\midrule
1.992 & 0.960119 & 0.00645212 & 0.968337 & 0.00469370  \\
1.994 & 0.958908 & 0.00605522 & 0.967138 & 0.00435104  \\
1.996 & 0.957654 & 0.00567948 & 0.965896 & 0.00403027 \\
1.998 & 0.956388 & 0.00532542  & 0.964616 & 0.00373069 \\
2.002  & 0.953750 & 0.00467471 & 0.961962 & 0.00318916\\
2.004  & 0.952383 & 0.00437703 & 0.960594 & 0.00294573\\
2.006  & 0.951017 & 0.00409665 & 0.959188 & 0.00271904\\
2.008  & 0.949616 & 0.00383241 & 0.957789 & 0.00250835\\
\bottomrule
\end{tabular}
\caption{Observables $n_{\sca_{0.05}}$ and $r_{0.002}$ respect to the value reported by \textit{Planck} 2018 results for the power law Starobinsky inflationary model    calculated  numerically for  values of the parameter $\beta$ close to $\beta=2$, considering $N_e=50$, and $N_e=60$.}
\label{observables_nSrplS_additional}
\end{center}
\end{table}

The initial conditions  in terms of cosmic time and conformal time  used for the numerical calculation of $A_{\sca_{0.05}}$ for $N_e=50$ and $N_e=60$ in the power--law Starobinsky inflationary model, are shown in Table \ref{plS_initial_ASNe50} and Table \ref{plS_initial_ASNe60},

\begin{table}[htbp]
\centering
\begin{tabular}{ccccc}
\toprule
 & \multicolumn{2}{c}{$N_e=50$} \\
\cmidrule(lr){2-3}
$\beta$ 
& $t_{\ini_{0.05}}$ & $\eta_{\ini_{0.05}}$    \\
\midrule
1.90 & $23051.3$ & $ -35856.8$   \\
1.95    & $71033$ & $ -60726.2$   \\
2.00   & $240347$ & $ -126360.6$&  \\
2.05   & $746652$ & $ -277666.8$&   \\
2.1 & $2.26742 \times 10^6$ & $-642670.8$ &  \\
\bottomrule
\end{tabular}
\caption{Initial conditions to the numerical calculation of $A_{\sca_{0.05}}$ considering $N_e=50$ for the power--law Starobinsky inflationary model.}
\label{plS_initial_ASNe50}
\end{table}

\begin{table}[htbp]
\centering
\begin{tabular}{ccccc}
\toprule
 & \multicolumn{2}{c}{$N_e=60$} \\
\cmidrule(lr){2-3}
$\beta$ 
& $t_{\ini_{0.05}}$ & $\eta_{\ini_{0.05}}$    \\
\midrule
1.90 & $24013.5$ & $ -36439.1$   \\
1.95    & $82152.6$ & $  -65757.1$   \\
2.00   & $325197$ & $-154587.9$&  \\
2.05   & $1.27543 \times 10^6$ & $ -413225.0$&   \\
2.1  & $4.98572 \times 10^6$ & $-1.20051 \times 10^6$ &  \\
\bottomrule
\end{tabular}
\caption{Initial conditions to the numerical calculation of $A_{\sca_{0.05}}$ considering $N_e=60$ for the power--law Starobinsky inflationary model.}
\label{plS_initial_ASNe60}
\end{table}

\subsection{Power law $\alpha$--Starobinsky Inflationary Model}

In this subsection,  our results for the power-law $\alpha$ -- Starobinsky inflationary model are presented. The parameter choices for $\alpha$ and $\beta$ are those reported by Saini and Nautiyal \cite{saini:2026a}, namely
$\{\alpha,\beta\}=\{(0.34,1.99),\ (1,2),\ (5,1.98),\ (10,2.05),\\
\ (20,1.90)\}$.

\begin{figure}[th!]
\centering
\includegraphics[scale=0.5]{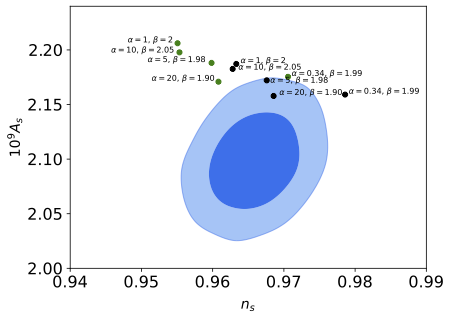}
\caption{The contour plot $(n_\sca,A_\sca,)$  for the power--law $\alpha$--Starobinsky inflationary model for  $N_e=50$ (green points) and $N_e=60$  (black points). In this Figure it can be observed that all points for $N_e=50$ are far from the Planck contour, and for $N_e=60$ the points $\alpha=\left\{(0.34,1.99),(1,2), (10,2.05) \right\}$ have the same behavior.}
\label{contour_ASnSplaS}
\end{figure}

\begin{figure}[th!]
\centering
\includegraphics[scale=0.5]{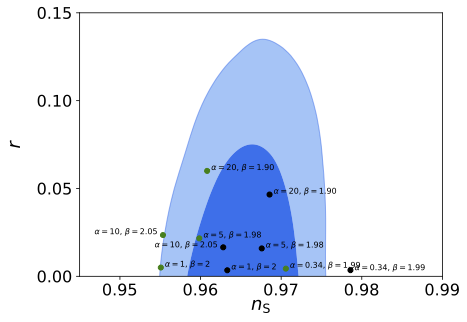}
\caption{The contour plot $(n_\sca,r)$ for the power--law $\alpha$--Starobinsky inflationary model $N_e=50$ (green points) and $N_e=60$  (black points). In this Figure it can be observed the point $\alpha=\left\{(0.34,1.99) \right\}$  for $N_e=60$ lives outside the Planck contour.}
\label{contour_nSrplaS}
\end{figure}

Figure \ref{contour_ASnSplaS} displays the $(n_\sca,A_\sca,)$ parameter space for $N_e=50$ and $N_e=60$, consistent with the value reported by Saini and Nautiyal \cite{saini:2026a}. We find that two of the points computed for $N_e=60$ fall within the $95\,\%$ confidence level of the \textit{Planck} constraints. Figure \ref{contour_nSrplaS} shows the $(n_\sca, r)$ plane for the same values of $N_e$. In this case, nearly all points obtained for $N_e=50$ and $N_e=60$ lie within the \textit{Planck} $68\,\%$ and $95\,\%$ confidence level, respectively, indicating that the model is strongly consistent with current observational bounds. The numerical values corresponding to these points are summarized in Tables \ref{observables_ASnSplaS} and \ref{observables_nSrplaS}.

\begin{table}[th!]
\begin{center}
\begin{tabular}{ccccc}
\toprule
 & \multicolumn{2}{c}{$N_e=50$} & \multicolumn{2}{c}{$N_e=60$} \\
\cmidrule(lr){2-3}\cmidrule(lr){4-5}
$\alpha$, $\beta$ 
& $n_{\sca_{0.05}}$ & $10^9 A_{\sca_{0.05}}$  
& $n_{\sca_{0.05}}$ & $10^9 A_{\sca_{0.05}}$  \\
\midrule
0.34, 1.99 & 0.970580 & 2.17539 & 0.978594 & 2.15911  \\
1, 2 & 0.955072 & 2.20613 & 0.963314 & 2.18717  \\
5, 1.98 & 0.959846 & 2.18813 & 0.967590 & 2.17215 \\
10, 2.05 & 0.955340 & 2.19777  & 0.962813 & 2.18261 \\
20, 1.90  & 0.960814 & 2.17089 & 0.968560 & 2.15785\\
\bottomrule
\end{tabular}
\caption{Observables  $n_{\sca_{0.05}}$ and $10^9 A_{\sca_{0.05}}$ respect to the value reported by \textit{Planck} 2018 results for the power--law $\alpha$--Starobinsky inflationary model  calculated  numerically for five different values of the parameter $\alpha$  and $\beta$ considering $N_e=50$, and $N_e=60$.}
\label{observables_ASnSplaS}
\end{center}
\end{table}

\begin{table}[th!]
\begin{center}
\begin{tabular}{ccccc}
\toprule
 & \multicolumn{2}{c}{$N_e=50$} & \multicolumn{2}{c}{$N_e=60$} \\
\cmidrule(lr){2-3}\cmidrule(lr){4-5}
$\alpha,\beta$ 
& $n_{\sca_{0.05}}$ & $r_{0.002}$ 
& $n_{\sca_{0.05}}$ & $r_{0.002}$ \\
\midrule
0.34, 1.99 & 0.970580 & 0.0043657 & 0.978594 & 0.0035044  \\
1, 2 & 0.955072 & 0.0049904 & 0.963314 & 0.0034505  \\
5, 1.98 & 0.959846 & 0.0216177 & 0.967590 & 0.0158912 \\
10, 2.05 & 0.955340 & 0.0234903  & 0.962813 & 0.0165201 \\
20, 1.90  & 0.960814 & 0.0600156 & 0.968560 & 0.0465538\\
\bottomrule
\end{tabular}
\caption{Observables $n_{\sca_{0.05}}$ and $r_{0.002}$ respect to the value reported by \textit{Planck} 2018 results for the power--law $\alpha$--Starobinsky inflationary model    calculated  numerically for five different values of the parameter $\alpha$ and $\beta$ considering $N_e=50$, and $N_e=60$.}
\label{observables_nSrplaS}
\end{center}
\end{table}

Results with $N_{e} =50$ also show consistency with those from Saini and Nautiyal \cite{saini:2026a}, where it is found, using MCMC, that parameters should rely on $\beta = 1.969^{+0.020}_{-0.023}$ and $\log_{10}\alpha = 0.37^{+0.82}_{-0.85}$ for $95\,\%$ C.L. where the analysis is combined with LSS observations. 

Results for $N_{e} = 60$ show good consistency with Planck 2018 contours, in contrast to the work in \cite{saini:2026a}, where these realizations are less favored. Additionally, this work reports derived confidence regions for $n_\sca$ and $r$ that are more constrained than the ones used here. However, our results show that most of the $N_e=60$ realizations for this model are consistent with these constrained contours: $r  = 0.019^{+0.015}_{-0.014}$ and $n_\sca = 0.9685_{-0.006}^{+0.0064}$, for $95\%$ confidence levels \cite{saini:2026a}, for the given selection of parameters $\alpha$ and $\beta$, except for the last one in Table \ref{observables_nSrplaS} : $\{\alpha, \beta\} = \{20, 1.90\}$, given its value of $r$.

The initial conditions  in terms of cosmic time and conformal time  used for the numerical calculation of $A_{\sca_{0.05}}$ for $N_e=50$ and $N_e=60$ in the power--law $\alpha$--Starobinsky inflationary model, are shown in Table \ref{plaS_initial_ASNe50} and Table \ref{plaS_initial_ASNe60},

\begin{table}[htbp]
\centering
\begin{tabular}{ccccc}
\toprule
 & \multicolumn{2}{c}{$N_e=50$} \\
\cmidrule(lr){2-3}
$\alpha$, $\beta$  
& $t_{\ini_{0.05}}$ & $\eta_{\ini_{0.05}}$    \\
\midrule
0.34, 1.99 & $279212$ & $ -139523.1$   \\
1, 2    & $240389$ & $  -126374.2$   \\
5, 1.98   & $72045.4$ & $ -61193.2$&  \\
10, 2.05   & $65444.4$ & $ -58126.0$&   \\
20, 1.90 & $24346.8$ & $-36642.6$ &  \\
\bottomrule
\end{tabular}
\caption{Initial conditions to the numerical calculation of $A_{\sca_{0.05}}$ considering $N_e=50$ for the power--law $\alpha$--Starobinsky inflationary model.}
\label{plaS_initial_ASNe50}
\end{table}

\begin{table}[htbp]
\centering
\begin{tabular}{ccccc}
\toprule
 & \multicolumn{2}{c}{$N_e=60$} \\
\cmidrule(lr){2-3}
$\alpha$, $\beta$ 
& $t_{\ini_{0.05}}$ & $\eta_{\ini_{0.05}}$    \\
\midrule
0.34, 1.99 & $336541$ & $ -158227.3$   \\
1, 2    & $325197$ & $  -154587.9$   \\
5, 1.98   & $97782.1$ & $ -72557.0$&  \\
10, 2.05   & $92831$ & $ -70435.2$&   \\
20, 1.90 & $34222.5$ & $-42321.2$ &  \\
\bottomrule
\end{tabular}
\caption{Initial conditions to the numerical calculation of $A_{\sca_{0.05}}$ considering $N_e=60$ for the power--law $\alpha$--Starobinsky inflationary model.}
\label{plaS_initial_ASNe60}
\end{table}

\section{Conclusions}
\label{Conclusions}

In this work, we present a numerical study of three Starobinsky --type inflationary scenarios: the $\alpha$--Starobinsky model, the power--law Starobinsky model, and the power--law $\alpha$--Starobinsky model. Each framework reduces to the standard Starobinsky inflationary model for suitable choices of its parameters.

In order to confront the theoretical predictions with observations, we present the resulting values in the $(n_\sca, A_\sca)$ and $(n_\sca, r)$ planes through contour plots. Within the parameter ranges explored, and for $N_e=50$ and $N_e=60$, we find that both the $\alpha$--Starobinsky and the power--law Starobinsky models exhibit regions of parameter space consistent with the \textit{Planck} 2018 constraints, making them observationally favored in those regimes. In the power--law Starobinsky case, values close to $\beta=2$ provide the best agreement with the data. Moreover, the power--law $\alpha$--Starobinsky scenario is more tightly constrained, it remains favored for the parameter choices considered in our analysis,  in accordance with the conclusion of Saini and Nautiyal \cite{saini:2026a} obtained in their Bayesian analysis. Overall, these scenarios can be regarded as well--motivated extensions of the Starobinsky inflationary model, capable of satisfying current \textit{Planck} bounds and opening the possibility of fitting with other observational data.

\section{Acknowledgments}
JZ is supported by the Doctoral Scholarship of the School of Postgraduate Studies and Continuing Education (EPEC Doctorate Scholarship) from FCFM.


\begin{thebibliography}{10}

\bibitem{guth:1981}
A.~H. Guth.
\newblock {Inflationary universe: A possible solution to the horizon and
  flatness problems}.
\newblock {\em Phys. Rev. D}, 23:347, 1981.

\bibitem{linde:1982}
A.~D. Linde.
\newblock {A new inflationary universe scenario: a possible solution of the
  horizon, flatness, homogeneity, isotropy and primordial monopole problems}.
\newblock {\em Phys. Lett. B}, 108:389, 1982.

\bibitem{guth:1985}
A.~H. Guth and S.~Pi.
\newblock {Quantum mechanics of the scalar field in the new inflationary
  universe}.
\newblock {\em Phys. Rev. D}, 32:1899, 1985.

\bibitem{martin:2014}
{J. Martin, C. Ringeval, and V. Vennin}.
\newblock {Encyclopaedia Inflationaris}.
\newblock {\em Phys. Dark Univ.}, 5--6:75, 2014.

\bibitem{mukhanov:1981}
V.~F. Mukhanov and G.V. Chibisov.
\newblock {Quantum fluctuations and a nonsingular universe}.
\newblock {\em Pis’ma Zh. Eksp. Teor. Fiz.}, 33:549, 1981.

\bibitem{jarosik:2011}
{N. Jarosik \textit{et al.}}
\newblock {Seven-year wilkinson microwave anisotropy probe (WMAP*)
  observations: Sky maps, systematic errors, and basic results}.
\newblock {\em ApJS}, 192:14, 2011.

\bibitem{akrami:2020}
Y.~Akrami \textit{et al.}
\newblock {Planck 2018 results. X. Constraints on inflation}.
\newblock {\em Astron. Astrophys.}, 641:A10, 2020.

\bibitem{starobinsky:1980}
{A. A. Starobinsky}.
\newblock {A new type of isotropic cosmological models without singularity}.
\newblock {\em Phys. Lett. B}, 91:99, 1980.

\bibitem{saini:2026a}
S.~Saini and A.~Nautiyal.
\newblock Power law $\ensuremath{\alpha}$ -starobinsky inflation.
\newblock {\em Phys. Rev. D}, 113:043546, 2026.

\bibitem{saha:2025}
Bhargabi Saha and Malay~K. Nandy.
\newblock {The $\alpha$-Attractor {E}-Model in Warm Inflation: Observational
  Viability from {Planck} 2018}, 2025.

\bibitem{saini:2023a}
{S. Saini and A. Nautiyal}.
\newblock {Observational constraints on power law Starobinsky inflation}.
\newblock {\em Phys. Rev. D}, 108:123505, 2023.

\bibitem{odintsov:2023}
S.~D. Odintsov and V.~K. Oikonomou.
\newblock {Generalized {$R^p$}-attractor Cosmology in the Jordan and Einstein
  Frames: New Type of Attractors and Revisiting Standard Jordan Frame {$R^p$}
  Inflation}.
\newblock {\em Int. J. Mod. Phys. D}, 32:2250135, 2023.

\bibitem{rodrigues:2023}
G.~Rodrigues-da Silva and L.~G. Medeiros.
\newblock {Second--order corrections to Starobinsky inflation}.
\newblock {\em EPJC}, 83:1032, 2023.

\bibitem{meza:2021}
{S. Meza, D. Altamirano, M. Z. Mughal, and Clara Rojas}.
\newblock {Numerical analysis of the generalized Starobinsky inflationary
  model}.
\newblock {\em Int. J. Mod. Phys. D}, 30:2150062, 2021.

\bibitem{canko:2020}
D.~D. Canko, I.~D. Gialamas, and G.~P. Kodaxis.
\newblock {A simple $F(\mathcal{R},\phi)$ deformation of Starobinsky
  inflationary model}.
\newblock {\em EPJC}, 80:458, 2020.

\bibitem{chakravarty:2015}
{G. K. Chakravarty and S. Mohanty}.
\newblock {Power law Starobinsky model of inflation from no--scale SUGRA}.
\newblock {\em Phys. Lett. B}, 746:242, 2015.

\bibitem{kallosh:2013b}
R.~Kallosh and A.~Linde.
\newblock {Universality Class in Conformal Inflation}.
\newblock {\em JCAP}, 07:002, 2013.

\bibitem{kallosh:2013a}
R.~Kallosh, A.~Linde, and D.~Roest.
\newblock {Superconformal Inflationary $\alpha$-Attractors}.
\newblock {\em JHEP}, 11:198, 2013.

\bibitem{ellis:2013}
J.~Ellis, D.~V. Nanopoulos, and K.~A. Olive.
\newblock {No-Scale Supergravity Realization of the Starobinsky Model of
  Inflation}.
\newblock {\em Phys. Rev. Lett.}, 111:111301, 2013.

\bibitem{roest:2014}
D.~Roest.
\newblock {Universality classes of inflation}.
\newblock {\em JCAP}, 01:007, 2014.

\bibitem{cecotti:2014}
S.~Cecotti and R.~Kallosh.
\newblock {Cosmological Attractor Models and Higher Curvature Supergravity}.
\newblock {\em JHEP}, 05:114, 2014.

\bibitem{noriji:2017}
Shin’ichi Nojiri, Sergei~D. Odintsov, and Vasilis~K. Oikonomou.
\newblock Modified gravity theories on a nutshell: Inflation, bounce and
  late--time evolution.
\newblock {\em arXiv: General Relativity and Quantum Cosmology}, 2017.

\bibitem{saini:2025a}
{S. Saini and A. Nautiyal}.
\newblock {Observational Constraints on $\alpha$--Starobinsky Inflation}.
\newblock {\em Gravitation \& Cosmology}, 31:517, 2025.

\bibitem{kehagias:2014}
{A. Kehagias, A. M. Dizgah, and A. Riotto}.
\newblock {Remarks on the Starobinsky model of inflation and its descendants}.
\newblock {\em Phys. Rev. D}, 89:043527, 2014.

\bibitem{ferrara}
{S. Ferrara, R. Kallosh, A. Linde, and M. Porrati}.
\newblock {Minimal supergravity models of inflation}.
\newblock {\em Phys. Rev. D}, 88:085038, 2013.

\bibitem{odintsov:2016}
S.D Odintsov and V.~K. Oikonomou.
\newblock {Inflationary $\alpha$-attractors from $F(R)$ gravity}.
\newblock {\em Phys. Rev. D}, 94:124026, 2016.

\bibitem{frolovsky:2025}
D.~Frolovsky and S.~V. Ketov.
\newblock {One-loop corrections to the {E}-type $\alpha$-attractor models of
  inflation and primordial black hole production}.
\newblock {\em Phys. Rev. D}, 111:083533, 2025.

\bibitem{ellis:2025}
J.~Ellis, M.~A.~G. Garcia, K.~A. Olive, and S.~Verner.
\newblock {Constraints on Attractor Models of Inflation and Reheating from
  {Planck}, {BICEP/Keck}, {ACT DR6}, and {SPT-3G} Data}.
\newblock {\em 2510.18656}, 2025.

\bibitem{alho:2017}
A.~Alho and C.~Uggla.
\newblock {Inflationary $\alpha$-attractor cosmology: A global dynamical
  systems perspective}.
\newblock {\em Phys. Rev. D}, 95:083517, 2017.

\bibitem{maeda:1989}
K.~I. Maeda.
\newblock {Towards the Einstein-Hilbert action via conformal transformation}.
\newblock {\em Phys. Rev. D}, 39:3159, 1989.

\bibitem{noriji:2011}
Shin'ichi Nojiri and Sergei~D. Odintsov.
\newblock {Unified cosmic history in modified gravity: From F(R) theory to
  Lorentz non--invariant models}.
\newblock {\em Phys. Rep.}, 505:59, 2011.

\bibitem{muller:1990}
{V. Muller, H. J. Schmidt, and A. A. Starobinsky}.
\newblock Power-law inflation as an attractor solution for inhomogeneous
  cosmological models.
\newblock {\em Class. Quantum Grav.}, 7:1163, 1990.

\bibitem{sebastiani:2014}
L.~Sebastiani, G.~Cognola, R.~Myrzakulov, S.~D. Odintsov, and S.~Zerbini.
\newblock Nearly starobinsky inflation from modified gravity.
\newblock {\em Phys. Rev. D}, 89, 2014.

\bibitem{cai:2014}
J.~O.~Gong Y.~F.~Cai and S.~Pi.
\newblock {Inflation beyond T-models and primordial B-modes}.
\newblock {\em Phys. Lett. B}, 738:20--24, 2014.

\bibitem{liddle2000cosmological}
{A. R. Liddle, and D. H. Lyth}.
\newblock {\em {Cosmological inflation and large--scale structure}}.
\newblock Cambridge university press, 2000.

\bibitem{Mathematica:2026}
{Wolfram Research, Inc.}
\newblock Mathematica, version 14.3.0.0.
\newblock https://www.wolfram.com/mathematica/, 2025.

\bibitem{mukhanov:1992}
V.~F. Mukhanov, H.~A. Feldman, and R.~H. Brandenberger.
\newblock {Theory of Cosmological Perturbations}.
\newblock {\em Phys. Rep.}, 215:203, 1992.

\bibitem{odintsov:2025a}
{S. D. Odintsov, and V. K. Oikonomou}.
\newblock {A power--law inflation tail for the standard $R^2$--inflation and
  the Trans--Planckian censorship conjecture}.
\newblock {\em Phys. Lett. B}, 865:139458, 2025.

\bibitem{odintsov:2025b}
{S. D. Odintsov, V. K. Oikonomou, E. I. Manouri, and A. T. Papadopoulos}.
\newblock {Solving the trans--Planckian censorship problem with a power--law
  tail in $R^2$ inflation: A dynamical system approach}.
\newblock {\em Nucl. Phys. B}, 1020:117114, 2025.

\bibitem{martin:2013}
J.~Martin and R.~H. Brandenberger.
\newblock {Trans--Planckian issues for inflationary cosmology}.
\newblock {\em Class. Quantum Grav.}, 30:113001, 2013.

\bibitem{habib:2005b}
{S. Habib, A. Heinen, K. Heitmann and G. Jungman}.
\newblock {Inflationary Perturbations and Precision Cosmology}.
\newblock {\em Phys. Rev. D}, 71:043518, 2005.

\bibitem{zambrano:2026}
{J. Zambrano, M. Agama, M. Garz\'on, W. Br\"amer--Escamilla, C. Rojas, and
  Te\'ofilo Vargas}.
\newblock {Quintessential inflation studied through semiclassical methods}.
\newblock {\em Int. J. Mod. Phys. D}, 35:2550993, 2026.

\bibitem{giare:2023b}
{W. Giar\`e, S. Pan, E. Di Valentino, W. Yang, J. de Haro, and A. Melchiorri}.
\newblock {Inflationary Potential as seen from Different Angles: Model
  Compatibility from Multiple CMB Missions}.
\newblock {\em JCAP}, 09:019, 2023.

\bibitem{das:2023}
{S. Das and R. O. Ramos}.
\newblock {Running and Running of the Running of the Scalar Spectral Index in
  Warm Inflation}.
\newblock {\em Universe}, 9:76, 2023.

\bibitem{vazquez:2020}
{J. A. Vazquez, L. E. Padilla, and T. Matos}.
\newblock {Inflationary cosmology: from theory to observations}.
\newblock {\em Rev. Mex. Fis. E}, 17:73, 2020.

\bibitem{finelli:2018}
{F. Finelli {\textit et al.}}
\newblock {Exploring cosmic origins with CORE: Inflation}.
\newblock {\em JCAP}, 04:016, 2018.

\bibitem{vazquez:2013}
{J. A. Vazquez, M. Bridges, Y--Z. Ma, and M. P. Hobson}.
\newblock {Constraints on the tensor--to--scalar ratio for non--power--law
  models}.
\newblock {\em JCAP}, 08:001, 2013.

\bibitem{haque:2025}
Md~Riajul Haque, Sourav Pal, and Debarun Paul.
\newblock {{ACT DR6} Insights on the Inflationary Attractor models and
  Reheating}, 2025.

\bibitem{burikham:2024}
P.~Burikham, T.~Chantavat, and P.~Boonaom.
\newblock Observational constraints on extended starobinsky and weyl gravity
  model of inflation.
\newblock {\em J. High Energy Astrophys.}, 43:178, 2024.

\bibitem{codello:2015}
A.~Codello, J.~Joergensen, F.~Sannino, and O.~Svendsen.
\newblock Marginally deformed starobinsky gravity.
\newblock {\em JHEP}, 2015:050, 2015.

\bibitem{ben:2014}
I.~Ben-Dayan, S.~Jing, M.~Torabian, A.~Westphal, and L.~Zarate.
\newblock {$R^2\log R$ quantum corrections and the inflationary observables}.
\newblock {\em JCAP}, 2014:005, 2014.

\bibitem{rodrigues2021observational}
{J. G. Rodr\'iguez, S. Santos da Costa, and J. S. Alcaniz}.
\newblock {Observational constraints on $\alpha$--attractor inflationary models
  with a Higgs--like potential}.
\newblock {\em Phys. Lett. B}, 815:136156, 2021.

\bibitem{adame:2025}
{A.G. Adame \textit{et al.}}
\newblock {DESI 2024 III: baryon acoustic oscillations from galaxies and
  quasars}.
\newblock {\em JCAP}, 04:012, 2025.

\bibitem{oikonomou:2025}
{V. K. Oikonomou}.
\newblock {Strong gravity eﬀects on $\mathcal{R}^2$--corrected single scalar
  field inflation and compatibility with the ACT data}.
\newblock {\em Phys. Lett. B}, 871:139972, 2025.

\bibitem{odintsov:2025c}
{S. D. Odintsov, and V. K. Oikonomou}.
\newblock {Power--law $F(R)$ gravity as deformations to Starobinsky inflation
  in view of ACT}.
\newblock {\em Phys. Lett. B}, 870:139907, 2025.

\end{thebibliography}

\end{document}